# Acoustic and Electromagnetic Co-Modeling of Piezoelectric Devices at Millimeter Wave

Tianyi Zhang*, Yen-Wei Chang*, Omar Barrera, Naveed Ahmed, Jack Kramer, and Ruochen Lu

*Abstract*— This work reports the procedure for modeling piezoelectric acoustic resonators and filters at millimeter wave (mmWave). Different from conventional methods for lower frequency piezoelectric devices, we include both acoustic and electromagnetic (EM) effects, e.g., self-inductance, in both the circuit-level fitting and finite element analysis, toward higher accuracy at higher frequencies. To validate the method, thin-film lithium niobate (LiNbO$_3$) first-order antisymmetric (A1) mode devices are used as the testbed, achieving great agreement for both the standalone resonators and a fifth-order ladder filter. Upon further development, the reported acoustic and EM co-modeling could guide the future design of compact piezoelectric devices at mmWave and beyond.

*Index Terms*— acoustic filters, device modeling, lithium niobate, millimeter-wave, piezoelectric devices, thin-film devices

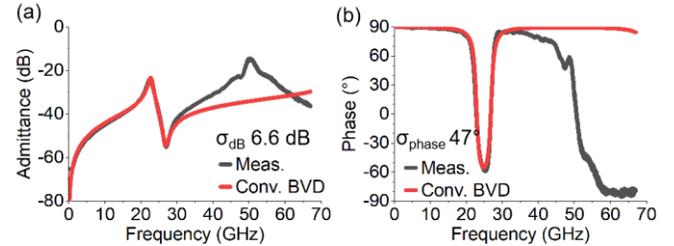

Fig. 1. Example measured mmWave acoustic resonator admittance in (a) amplitude and (b) phase, compared with that fitted with conventional BVD model.

## I. Introduction

PIEZOELECTRIC devices at radio frequencies (RF) have been commercially used as the dominant sub-6 GHz front-end filter solution in mobile devices [1]–[5]. The main advantage of acoustics over conventional electromagnetic (EM) elements is their compact size, marked by four to five orders of magnitude smaller wavelengths, which allows reduced physical sizes suitable for mobile applications [1], [6], [7]. Recently, due to increasing demands for wireless bandwidth, RF front-end devices are evolving towards higher frequencies, focusing on millimeter wave (mmWave) bands. Acoustic devices could play a similar role for mmWave front-ends, if they can be frequency scaled into mmWave without significantly losing performance. However, such frequency scaling is not trivial, as the reduction in acoustic wavelength introduces additional fabrication and design difficulties compared to lower frequencies. Further development calls for innovations in both the material [8], [9], microfabrication [10]–[12], and device design [13]–[15].

More recently, researchers have achieved a notable level of success in demonstrating acoustic resonators in piezoelectric thin films [16], [17], by leveraging thickness modes, e.g., first-order symmetric (S1) and antisymmetric (A1) modes. So far, promising material candidates include aluminum nitride/scandium aluminum nitride (AlN/ScAlN) [13], [14], [16]–[21] and lithium niobate (LiNbO$_3$) [11], [22]–[25]. Through the development of these device platforms, resonators with improved figure-of-merit (FoM), which is the product of quality factor ($Q$) and electromechanical coupling coefficient ($k^2$), have been achieved at mmWave. Built upon standalone resonators, mmWave acoustic filters have been successfully developed [26]–[28]. Despite the promising results from the prototypes, further design and optimization are still required to meet the requirements of real-world wireless networks.

One major design bottleneck is accurately modeling acoustic resonators and filters at mmWave. One example of measured admittance of mmWave acoustic resonators is shown in Fig. 1 (a) and 1(b) in amplitude and phase, respectively. The EM self-resonance can be clearly identified, marked by the inductive admittance phase, which a multi-branch BVD model cannot capture. This EM self-resonance arises from a combination of the inductance in transducers and routing, along with the naturally capacitive response of acoustic device transducers. Unlike in the lower frequency case, where such effects could be ignored as they are far from the passband of interest, such EM effects cannot be neglected when extracting key device performance with Butterworth-Van Dyke (BVD) models or optimizing device performance. Unfortunately, current piezoelectric finite element analysis (FEA), e.g., COMSOL, does not capture the EM effects. Similarly, it is challenging to incorporate acoustic models in EM simulators, e.g., HFSS, mostly due to the dramatic wavelength difference between acoustic and EM domains. As the frequency scaling continues, the EM effects can be more impactful. Moreover, it would be challenging to design higher-order filters without considering the parasitic feedthrough and routing effects. Further development of mmWave piezoelectric devices calls for the capability to accurately co-model EM and acoustic effects.

In this work, we introduce an innovative modeling approach that utilizes acoustic and EM co-modeling. This approach

Manuscript received 1 June 2024; revised XX June 2024; accepted XX June 2024. This work was supported by DARPA COmpact Front-end Filters at the ElEment-level (COFFEE).

T. Zhang, Y.-W. Chang, O. Barrera, N. Ahmed, J. Kramer, and Ruochen Lu and R. Lu are with The University of Texas at Austin, Austin, TX, USA (email: yenwei.chang@utexas.edu). T. Zhang and Y.-W. Chang are co-first authors.

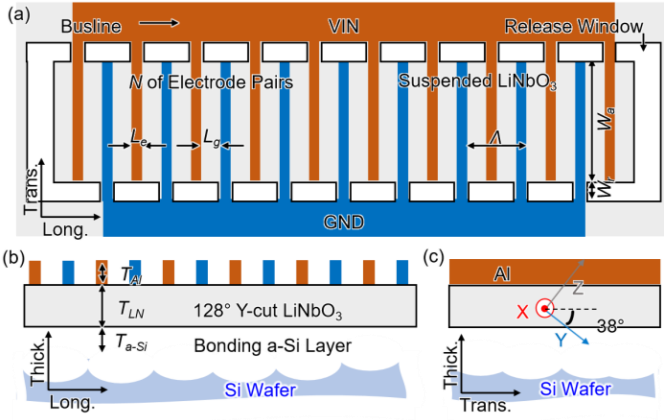

Fig. 2. Mockup of an A1 resonator in a suspended 128° Y-cut LiNbO$_3$ thin film. (a) Top view, (b) front view, and (c) side view.

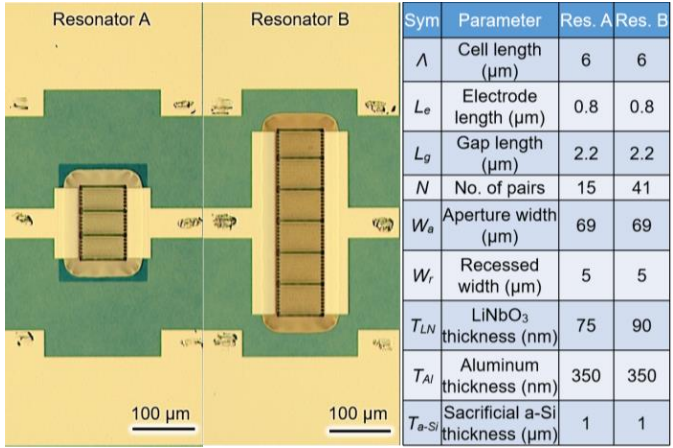

Fig. 3. Optical microscope images of fabricated resonators A and B. Key design dimensions are listed in the table.

integrates EM effects, such as self-inductance, into both circuit-level fitting and finite element analysis-based simulation, significantly improving mmWave acoustic resonator model accuracy. The effectiveness of this method has been validated using thin film LiNbO$_3$ first-order antisymmetric (A1) mode devices as a testbed, showing good agreement between the model and measurement of both standalone resonators and a synthesized fifth-order ladder filter. This new method could guide future designs of compact piezoelectric devices at mmWave and beyond.

## II. EQUIVALENT CIRCUIT MODEL

### A. Testbed Piezoelectric Resonators at mmWave

To validate our modeling approach, mmWave resonators, and corresponding filters [27] were used as testbeds. The resonator includes an interdigital transducer array on a 75-90 nm thick 128° Y-cut LiNbO$_3$ thin film over a silicon substrate with a 1 μm thick amorphous silicon bonding layer for ease of mechanical suspension. The design employs 350 nm thick aluminum electrodes to reduce electrical loss. The electric field generated between IDT fingers excites the A1 mode via piezoelectric coefficient $e_{15}$. The top, front, and side schematic of the resonators are shown in Fig. 2(a)-(c). Key design parameters are labeled in Fig. 2 and listed in the inset table in Fig. 3. The fabrication flow has been reported in [27]. The optical images of the resonator are illustrated in Fig. 3. Ground-signal-ground (GSG) probing pads with 200 μm spacings between tips are included as the two-port configuration. The electrodes are grouped to minimize the isotropic silicon etch release distance.

### B. Modified mmWave BVD Model

The modified mmWave BVD model we utilized comprises three components: the motional elements, static capacitance ($C_0$), and EM parasitic elements. The schematic of our MBVD model is shown in Fig. 4, which details the process for obtaining motional elements, namely motional resistance ($R_m$), motional capacitance ($C_m$), and motional inductance ($L_m$), calculated from resonant frequency ($f_s$), quality factor ($Q$), and coupling coefficient ($k^2$). The EM terms include the series resistance ($R_s$) and inductance ($L_s$) associated with electrodes and leading buslines, as well as feedthrough capacitance ($C_f$) between the ports and the ground. The model is similar to those reported in

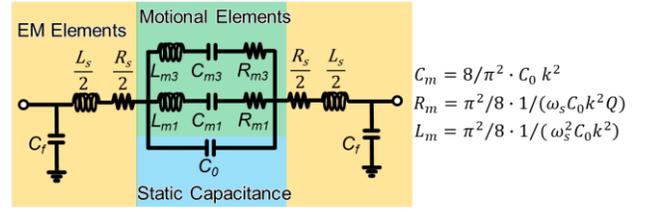

$$C_m = 8/\pi^2 \cdot C_0 k^2$$
$$R_m = \pi^2/8 \cdot 1/(\omega_s C_0 k^2 Q)$$
$$L_m = \pi^2/8 \cdot 1/(\omega_s^2 C_0 k^2)$$

| Sym | Param. | Resonator A | | Resonator B | |
|---|---|---|---|---|---|
| $C_0$ | Static capacitance (fF) | 50 | | 155 | |
| $L_s$ | Series inductance (pH) | 200 | | 150 | |
| $C_f$ | Feedthrough capacitance (fF) | 15 | | 15 | |
| $R_s$ | Series resistance (ohm) | 6 | | 2.8 | |
| **Motional Elements** | | A1 | A3 | A1 | A3 |
| $f_s$ | Series resonance (GHz) | 23.9 | N/A | 19.8 | 59.4 |
| $k^2$ | Electromechanical coupling (%) | 43 | N/A | 44 | 4.88 |
| $Q$ | Quality factor | 32 | N/A | 40 | 20 |
| $C_m$ | Motional capacitance (fF) | 17.42 | N/A | 55.28 | 6.13 |
| $R_m$ | Motional resistance (ohm) | 11.94 | N/A | 3.64 | 21.81 |
| $L_m$ | Motional inductance (pH) | 2544 | N/A | 1168 | 1171 |

Fig. 4. Modified mmWave BVD, along with key extracted parameters.

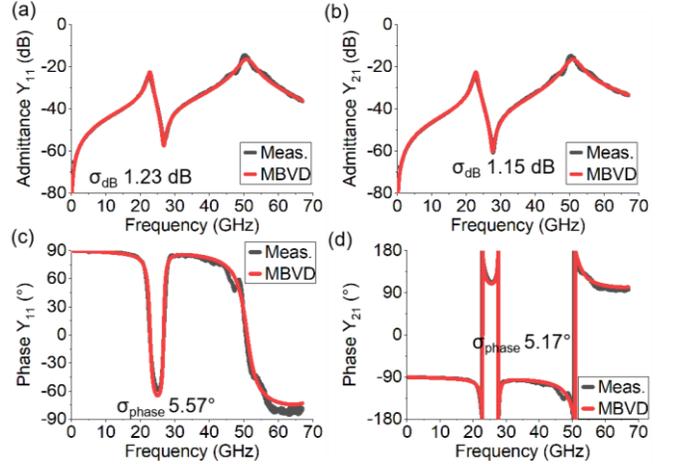

Fig. 5. Measured wideband admittance of resonator A (a) $Y_{11}$ and (b) $Y_{21}$ in amplitude, (c) $Y_{11}$ and (d) $Y_{21}$ in phase with modified mmWave BVD fitting.

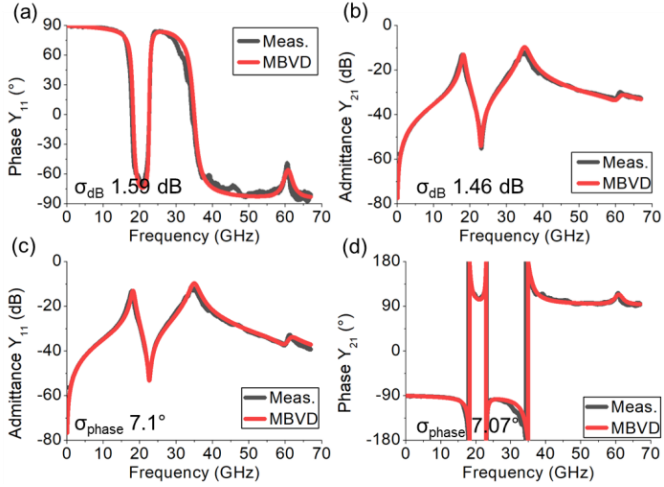

Fig. 6. Measured wideband admittance of resonator B (a) $Y_{11}$ and (b) $Y_{21}$ in amplitude, (c) $Y_{11}$ and (d) $Y_{21}$ in phase with modified mmWave BVD fitting.

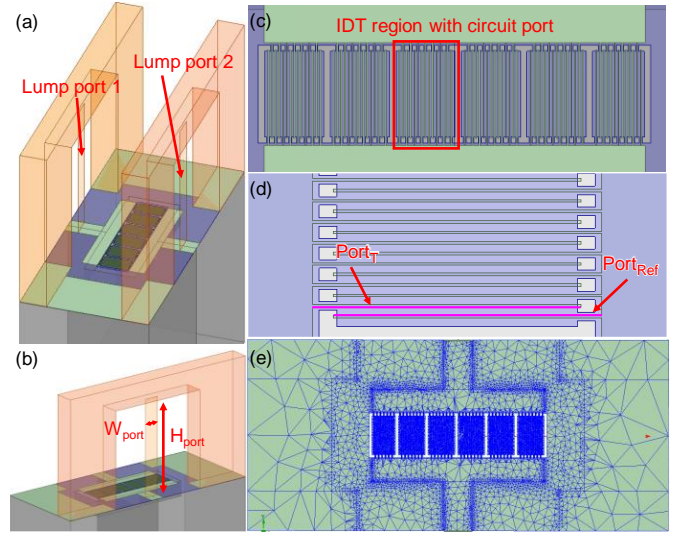

Fig. 7. (a) HFSS model setup using the example of the resonator B. (b) The port setup on one side. (c) IDTs region. (d) Zoom-in view of the IDTs region with circuit ports. (e) Mesh setting after including the circuit ports.

lower-frequency devices [29]. However, in this work, we look at the wideband impact, as the mechanical resonances and self-resonances are all around tens of GHz. This work includes a few examples of such effects and also provides a brief guide on the fitting strategy.

The measurement of the resonator A is presented in Fig. 5. For each device, we report admittance elements $Y_{11}$ and $Y_{21}$ in amplitude [Fig. 5(a) and 5(b)] and phase [Fig. 5(c) and 5(d)], while $Y_{22}$ and $Y_{12}$ are identical due to the device's symmetry and reciprocity. Unlike conventional piezoelectric resonators at lower frequencies, strong EM self-resonance, from $L_s$ and $C_0$, is seen above acoustic resonance, which will be accounted for by the modified mmWave BVD model. For admittance $Y_{21}$, the A1 resonance is at 23.9 GHz, while the self-resonance is at 50.5 GHz. The A3 resonance is expected to be 71.7 GHz, which is out of the range of the 67 GHz network analyzer.

The recursive fitting procedure is presented below. First, the EM resonance and off-resonance capacitance are used for extracting $L_s$, $C_0$, and $C_f$. $C_0$ and $C_f$ are extracted by admittance, Y-parameters, using the imaginary part magnitude. $L_s$ is derived through EM resonance frequency. Second, the motional elements are extracted after removing the effects of $L_s$ for A1 and A3 consequently. Finally, a recursive process is used toward convergence.

The example fittings of the resonator A are shown in Fig. 5(a)-(d) in both amplitude and phase in red. Both the first-order antisymmetric (A1) and third-order antisymmetric (A3) modes below 67 GHz (network analyzer maximum frequency range) are fitted, respectively. A good agreement is obtained. The extracted EM and acoustic parameters are listed in the inset table in Fig. 4. Similarly, for resonator B, the measurements of admittance parameters in amplitude and phase are plotted in Fig. 6(a)-(d), along with the fitting results. The extracted parameters are also listed in the inset table of Fig. 4.

To judge the quality of the fitting quantitatively, two standards, the standard deviation of the amplitude in the decibel scale $\sigma_{dB}$ and the standard deviation of the phase in the degree scale $\sigma_{phase}$ are defined as:

$$\sigma_{dB} = \sqrt{\Sigma\left(dB(Y_m) - dB(Y_f)\right)^2 / N} \quad (1)$$

$$\sigma_{phase} = \sqrt{\Sigma\left(\text{phase}(Y_m) - \text{phase}(Y_f)\right)^2 / N} \quad (2)$$

where $Y_m$ is the measurement admittance, $Y_f$ is the fitting admittance, N is the number of data points. The phase difference is wrapped within 90°. The fitting result of $Y_{11}$ of resonator A has a $\sigma_{dB}$ of 1.23 dB and a $\sigma_{phase}$ of 5.57°. In comparison, the standard deviation from Fig. 1, which is the fitting of $Y_{11}$ of resonator A using a conventional BVD model, exhibits a $\sigma_{dB}$ of 6.6 dB and a $\sigma_{phase}$ of 47°. The fitting could be further improved if more motional branches are included to include the spurious modes beyond A1 and A3.

The deviation of other parameters of both resonator A and resonator B are labeled in Fig. 5 and Fig. 6. For admittance $Y_{21}$, the A1 resonance is at 19.8 GHz, the A3 resonance is at 59.4 GHz, and the self-resonance is at 34.3 GHz. It is interesting to note that the resonator becomes less inductive near A3 resonance, different from lower frequency modes, due to the EM resonance. The great agreement validates the equivalent circuit model. The accuracy of the fitting can be further improved by adding distributed elements to represent the routings. However, for easier implementation, we are staying at lumped EM elements for this section but will use the FEA method for modeling with higher accuracy.

The comparison between $R_s$ and $L_s$ in resonators A and B is also noteworthy. Compared to Resonator A, Resonator B has more electrodes than A (2.7 times) and a longer busline (2.6 times). The longer busline contributes to larger $R_s$ and $L_s$ (longer trace), while more electrodes lead to smaller $R_s$ and $L_s$ (more electrodes in parallel). For our specific design, $R_s$ and $L_s$ of resonator B are smaller than A by 2.1 times and 1.3 times, respectively. This indicates that the electrodes have a significant impact, and that the busline seems to be more impactful to inductance. Further studies are needed on this end, while this paper focuses on the methodology for fitting and simulating acoustic resonators and filters at mmWave.

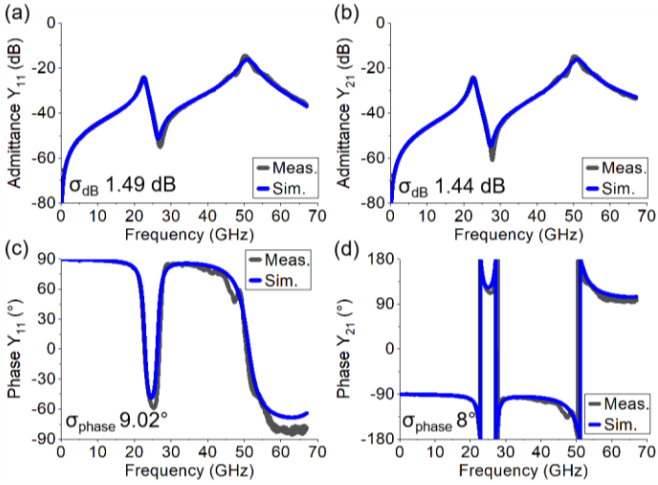

Fig. 8. FEA EM-acoustic co-modeled wideband admittance of resonator A (a) $Y_{11}$ and (b) $Y_{21}$ in amplitude, (c) $Y_{11}$ and (d) $Y_{21}$ in phase, plotted against measurement.

## III. FEA SIMULATION OF MMWAVE ACOUSTIC RESONATOR

The schematic representation of the resonator B in the EM FEA simulator HFSS is depicted in Fig. 7(a). The red component situated atop the resonator is incorporated to connect two distinct ground planes, emulating the probe configuration. Port assignments involved using lump ports at both ends for signal excitation. Lump port dimensions were matched to actual G-S-G probe sizes with a port width ($W_{port}$) of 50 µm and a port height ($H_{port}$) of 350 µm. The port setup on one side is plotted in Fig. 7(b). In addition, the material of the probe structure is set as a perfect electrical conductor to avoid inducing unwanted loss from the probes.

Two main issues remain for the process. First, properly incorporating motional elements $R_m$, $L_m$, and $C_m$ into co-modeling simulation is crucial. Second, properly selecting meshing is important. In our specific case, where we are simulating a device with a thin film of LiNbO$_3$ with a thickness of 75nm and a narrow interdigital transducer (IDT) structure, a fine mesh is necessary to ensure precise simulation $L_s$ and $R_s$. However, excessively fine meshing over the entire device region can lead to memory overflow issues and unnecessarily long computing time.

To resolve these two issues, we incorporate circuit ports positioned between two electrodes fingers. The motional branch could be added between two ports on two adjacent fingers of IDT: Port$_T$ and Port$_{Ref}$ which are illustrated in Fig. 7(c) and 7(d). Motional elements, simulated from COMSOL FEA, will be connected to the circuit ports later with Keysight Advance Design System. The simulation process is thoroughly explained in [27]. This facilitates the co-modeling of acoustic and EM. On the mesh side, the HFSS simulator's automatic adaptive meshing process enhances the mesh resolution surrounding these circuit ports, promoting model accuracy [Fig. 7(e)].

The EM-acoustic co-simulated models are compared against the measurement. Fig. 8 and Fig. 9 show the admittance and phase response comparison between simulation and measurement for resonator A and resonator B, respectively. The grey lines depict the measurements, and the blue lines illustrate simulation results. To judge the quality of the simulation, the standard deviation of the simulation result $\sigma_{dB}$ and $\sigma_{phase}$ are adapted, using Eqs. 1-2 with measured and simulated admittance. The deviation of the simulation results of both resonators A and B are labeled in Fig. 8 and Fig. 9. Additionally, we fit the measurement results and the HFSS simulation result with the MBVD model and extract the parameters' values, which are listed in the table shown in Fig. 9. The results validate the effectiveness of the proposed method. The minor discrepancy is around the high impedance part, namely the shunt resonance of A1, likely caused by the capacitive feedthrough between the HFSS ports. This could be further improved by port structure optimization and shifting of the reference plane with additional simulation on calibration structures.

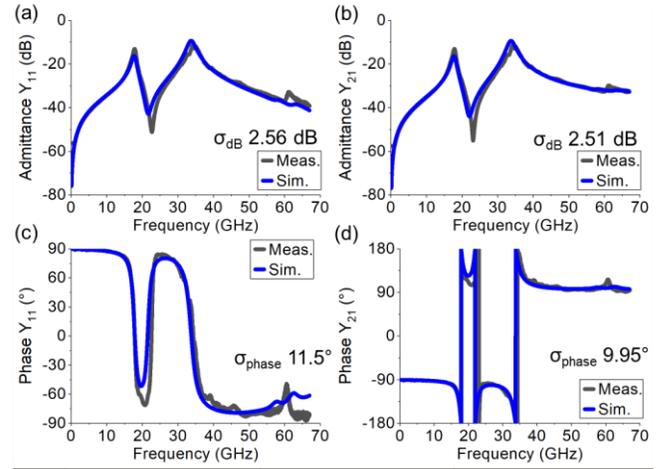

|  |  | Resonator A | | Resonator B | |
|---|---|---|---|---|---|
| Sym. | Param | Ext. from measurement | Ext. from HFSS | Ext. from measurement | Ext. from HFSS |
| $C_0$ | Static capacitance (fF) | 50 | 51.63 | 155 | 163.1 |
| $L_s$ | Series inductance (pH) | 200 | 192.37 | 150 | 151.47 |
| $C_f$ | Feedthrough capacitance (fF) | 15 | 17.18 | 20 | 27.4 |
| $R_s$ | Series resistance (ohm) | 6 | 5.8 | 2.8 | 2.76 |

Fig. 9. FEA EM-acoustic co-modeled wideband admittance of resonator B (a) $Y_{11}$ and (b) $Y_{21}$ in amplitude, (c) $Y_{11}$ and (d) $Y_{21}$ in phase, plotted against measurement. Extracted parameters are listed in the table.

This work so far focuses on a case-by-case basis. In general, acoustic resonators and filters at mmWave tend to suffer from significant EM effects. The main reason is that, despite the acoustic wavelengths, determined by the film thickness, being orders of magnitude smaller than that of EM wavelength, the lateral dimensions of the resonators are no longer electrically small (<0.1 wavelength [30]), due to the need for 50 Ω impedance matching. For instance, along the X-axis of 128° Y-cut LiNbO$_3$, the EM wavelength is 2.2 mm at 20 GHz and 0.9 mm at 50 GHz. Thus, devices with routing or electrodes longer than 220/90 µm are likely required to include the EM effects, e.g., examples in the work. Note that certain acoustic technologies with higher capacitance density, e.g., mmWave ScAlN FBARs [31]–[33], are more compact and thus might be less prone to EM effects. Nevertheless, considering EM effects

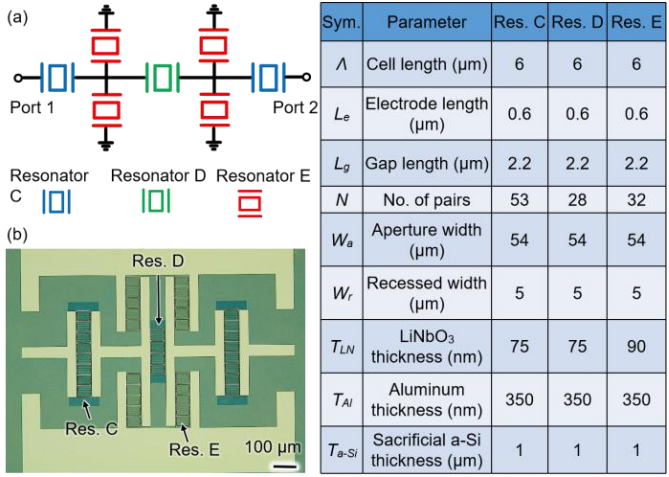

Fig. 10. Fifth-order ladder filter's (a) structure and (b) microscopic image.

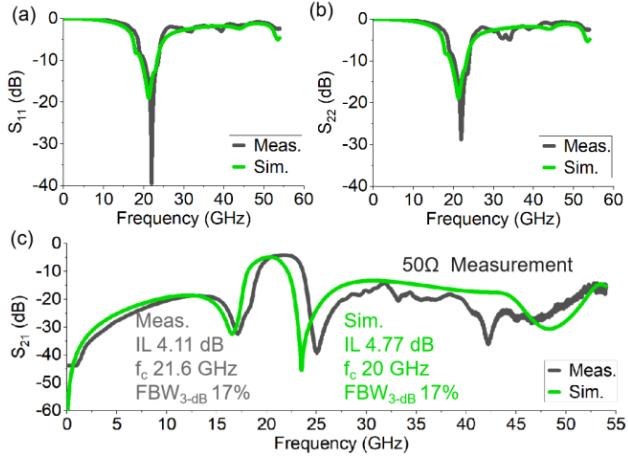

Fig. 11. Filter measurement in comparison with the simulated results from the proposed acoustic-EM co-modeling.

is still crucial for accurately simulating and modelling high-frequency acoustic technologies.

## IV. FEA SIMULATION OF MMWAVE ACOUSTIC FILTERS

After the validation of the modeling process on resonators, our focus extends toward system-level verification using filters. This is being undertaken through the development of a mmWave fifth-order ladder filter operating at 21.6 GHz, fabricated on a thin film $LiNbO_3$. The filter exhibits an insertion loss (IL) of 4.11 dB and a 3-dB fractional bandwidth (FBW) of 17%. This filter's structure includes two distinct types of series resonators [Resonator C and Resonator D in Fig. 10(b)] and four identical shunt resonators [Resonator E in Fig. 10(b)], as depicted in Fig. 10(a), accompanying a microscopic image [Fig. 10(b)], and the inset table shows the dimensions.

Fig. 11 shows a comparison between simulated and actual results of a fifth-order ladder filter, highlighting a strong correlation. The simulation shows an IL of 4.77 dB and a 3-dB FBW of 17% at 20 GHz. The deviations in the simulation can primarily be attributed to two factors: The first concerns the non-uniform thickness of the sample, which causes shifts in the resonances of Resonators C-E, consequently shifting the passband frequency. The second factor involves minor spurious modes, which are responsible for the observed subtle ripples above the passband. Overall, the results validate the proposed EM-acoustic co-model by applying it to a more complicated RF acoustic device.

Note that the current work focuses on presenting the methodology for fitting and simulating acoustic resonators and filters at mmWave. However, a full analysis of the topic is still far from being completed. For instance, one key topic is deriving estimated equations for inductance between the electrodes and along routings, following the examples in [34]. However, the major hurdle is the limited data of various acoustic technologies at mmWave. The authors are working toward this topic with effort on device design, microwave measurement, and theory aspects.

## V. CONCLUSIONS

This research outlines a novel co-modeling approach for piezoelectric devices at mmWave, integrating both acoustic and EM analyses. By employing a modified mmWave BVD model for fitting and optimizing HFSS port settings to incorporate acoustics in EM FEA, we maintain model accuracy. The methodology's effectiveness is validated by a good agreement between simulation and measurement in resonators and a synthesized fifth-order ladder filter. Overall, our methodology offers a viable solution for effectively predicting the responses of mmWave resonators and filters.

## ACKNOWLEDGMENT

The authors thank the DARPA COFFEE program for funding support and Dr. Ben Griffin, Dr. Todd Bauer, and Dr. Zachary Fishman for helpful discussions.